\documentclass[twocolumn,prl,showpacs,preprintnumbers,amsmath,amssymb]{revtex4}

\usepackage{graphics}
\usepackage{color}
\usepackage{epsfig}

\usepackage{graphicx}
\usepackage{dcolumn}
\usepackage{bm}

\begin{document}


\title{Association of Efimov trimers from a three-atom continuum}

\author{Olga Machtey, Zav Shotan, Noam Gross}
\author{Lev Khaykovich}%
 \email{lev.khaykovich@biu.ac.il}
\affiliation{%
Department of Physics, Bar-Ilan University, Ramat-Gan, 52900 Israel
}%


\begin{abstract}
We develop an experimental technique for rf-association of Efimov
trimers from a three-atom continuum. We apply it to probe the lowest
accessible Efimov energy level in bosonic lithium in the region
where strong deviations from the universal behavior are expected,
and provide a quantitative study of this effect. The position of the
Efimov resonance at the atom-dimer threshold, measured using a
different experimental technique, concurs with the rf-association
results.
\end{abstract}

\maketitle

Low energy few-body physics with resonantly enhanced two-body
interactions is a fundamental quantum mechanical problem, known to
manifest universal properties~\cite{Braaten&Hammer06}. For three
interacting particles universality is associated with a ladder of
weakly bound triatomic states called Efimov trimers that exist even
when the two-body interactions don't support any~\cite{Efimov70}. Recent
experiments with ultracold atoms have been highly successful in
demonstrating several aspects of
universality~\cite{Kraemer06,Knoop09,Zaccanti09,Pollack09,Gross09,
Barontini09,Ferlaino09,Ottenstein08,Huckans09,Williams09,Lompe10,Nakajima10,Gross10,
Berninger11,Lompe10RF,Nakajima11}.

In the limit of zero collision energy, the two-body interactions are
determined by a single parameter, the s-wave scattering length $a$,
which can be tuned to very large values near a Feshbach
resonance~\cite{Chin10}. The window of universality is opened when
$a$ greatly exceeds the characteristic range of the interatomic
potential $r_{0}$, basically equivalent to the van der Waals length
$r_{\textrm{vdW}}=\left(mC_{6}/16\hbar^2\right)^{1/4}$, where
$C_{6}$ is the van der Waals coefficient and $m$ is the atomic
mass~\cite{Gribakin93}. Analytical expressions for three-body
experimental observables are derived in this window in the zero
range limit ($r_{0}\rightarrow 0$) and their log-periodic dependence
on $a$ reveals the discrete scaling invariance of the underlying
Efimov physics~\cite{Braaten&Hammer06}. The impact of the finite
range of interaction potentials on the lowest Efimov energy level in
the context of ultracold atoms has been considered in recent
publications~\cite{Thogersen08,Platter09,DIncao09,Ji10,Naidon11,footnote01}
and its role in the ongoing experiments is the subject of an ongoing
debate~\cite{Knoop09,Zaccanti09,Pollack09,DIncao09,Ji10,Naidon11,Thogersen09}.
Corrections to the zero range approximation can be taken into
account via the effective range of the interatomic potential
$R_{e}$, which is extracted from the s-wave scattering phase shift
$\delta(k)$ at small relative wavenumbers $k$ using the effective
range expansion $k\cot\delta(k)=-1/a+R_{e}k^2/2$. Another
theoretical approach shows that deviations from universality can be
directly related to the Feshbach resonance
parameters~\cite{Jona-Lasinio10,Pricoupenko10,Pricoupenko11}.

Efimov physics in ultracold atoms is studied experimentally in the
vicinity of Feshbach resonances which allow $a$ to be tuned via the
external magnetic field bias~\cite{Chin10}. The usual three-body
observable which shows log-periodic dependence on $a$ constitutes
the three-body recombination loss rate ($K_{3}$) of atoms from a
shallow optical
trap~\cite{Kraemer06,Knoop09,Ferlaino09,Zaccanti09,Gross09,Pollack09,Barontini09,
Ottenstein08,Huckans09,Williams09,Lompe10,Nakajima10,Gross10,Berninger11}.
This experimental approach is only sensitive to a few points
associated with each Efimov energy level. For negative scattering
length, enhancements in $K_{3}$ around
certain values of $a$  enable one to probe the Efimov resonances
which occur when the trimer states intersect with the three-atom
continuum threshold~\cite{Kraemer06}. For positive scattering
length, the two lowest Efimov potentials open two pathways to a
weakly bound dimer which interfere destructively for certain values
of $a$, resulting in the appearance of recombination minima in
$K_{3}$ spectrum~\cite{Kraemer06,Zaccanti09}. In addition, the
signatures of the Efimov resonances at the atom-dimer threshold can
be observed through secondary collision processes~\cite{Zaccanti09}.

\begin{figure}
{\centering \resizebox*{0.41\textwidth}{0.35\textheight}
{{\includegraphics{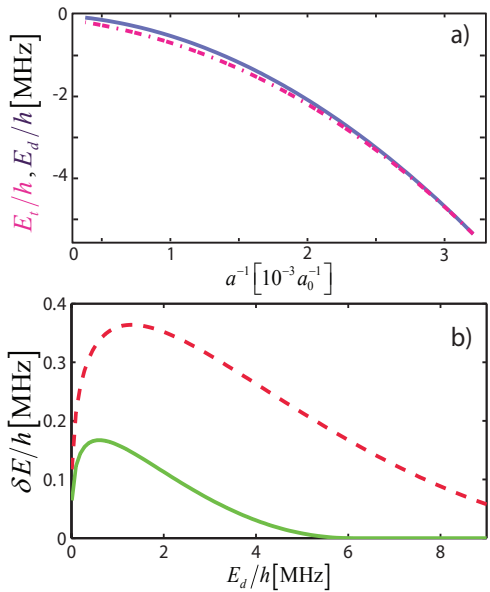}}}
\par}
\caption{\label{F1-EfimovScenario} (a) Dimer (solid blue) and trimer
(dotted-dashed pink) energy levels follow each other very closely as
calculated using the universal theory~\cite{Braaten&Hammer06} with
$a_{*}=288a_{0}$, where $a_{0}$ is the Bohr radius. This value
is expected from the previously measured three-body recombination
spectrum\cite{Gross10,Gross11}. (b) The difference between the dimer and
trimer energy levels for two different Efimov resonances at the
atom-dimer threshold: $a_{*}=288a_{0}$ (solid green) and $a_{*}=200a_{0}$
(dashed red).}
\end{figure}

Here we report on a different experimental technique in which we
form Efimov trimers directly from the three-atom continuum.
This approach applies a strong in-resonance rf-modulation of the
magnetic field in order to stimulate free-atoms to trimer transition in the
vicinity of the Efimov resonance at the atom-dimer threshold. At
this limit the trimer is strongly asymmetric and resembles a
composite dimer made up of a dimer and a nearly free
atom. Fig.~\ref{F1-EfimovScenario}(a) shows how similar the dimer and
the trimer energy levels are expected to be in this region. Near the
threshold point the atom-dimer scattering length is resonantly
enhanced~\cite{Braaten&Hammer06} and can greatly exceed $a$.
This gives the trimer wavefunction sufficient spatial extension to
overlap with the free three-atom scattering state. High rates of
dimer association were recently observed in our system in the
relevant range of $a$~\cite{Gross10,Gross11}. Here we extend this
approach and study the association of trimers. Note that the
lifetime of the trimers is expected to be short, since it will either dissociate
into high kinetic energy constituents (a deeply bound dimer and an
atom) or an atom-trimer inelastic collision will occur. Thus, trimers
disappear from the trap quickly and their binding energy shows up as a
loss resonance in rf-frequency scan measurements. This method
provides, as a function of $a$, a continuous spectroscopy probe of the
Efimov energy level and reveals its clear deviation from the
predictions of universal theory. We attribute this to the
manifestation of range corrections. The general trend of our data
conforms with the shifted position of the Efimov resonance at the
atom-dimer threshold that is detected through the secondary collision
resonance in the measurements of $K_{3}$.
Unfortunately, only a partial theory is currently available for
bosonic lithium, and it predicts a single point in the energy spectrum
- the position of the Efimov resonance at the atom-dimer
threshold~\cite{Ji10}. This prediction is in agreement
with our results.

A similar experimental approach has recently been explored for
fermionic $^6$Li atoms~\cite{Lompe10RF,Nakajima11}. It is important to note
two important differences. Primarily, extreme collisional stability
of weakly bound dimers formed by two fermions~\cite{Cubizolles03}
allows rf-association of trimers from the atom-dimer
continuum~\cite{Lompe10RF}. This property, being the result of the Pauli
exclusion principle, is not applicable in the case of bosonic $^7$Li and makes
the implementation of this experimental approach challenging. Therefore, we
explore the direct association of trimers from the three-atom
continuum which is more general and applicable to all particles,
independent of their quantum statistical nature. Secondly, the two
isotopes have very different Feshbach resonances,
conveniently classified by the resonance strength parameter
$s_{res}=2r_{0}/|R_{e}|$~\cite{Chin10}. The resonances in $^6$Li are
strongly entrance-channel dominated and thus characterized by
$s_{res}\gg 1$, corresponding to hundreds of Gauss of the magnetic field range where $R_{e}\ll r_{0}$. In contrast, the
resonances in $^7$Li, while being almost as broad as their $^6$Li
counterparts, are in between being
entrance-channel and close-channel dominated with $s_{res}\sim 1$.
This difference, caused by the anomalously large background
scattering length in $^6$Li, introduces the possibility of studying the
consequences of the early breakdown of universality in two-body physics
for the deviations from the universal behavior in three-body physics.

The experiments are performed on a gas of $^7$Li atoms polarized on
the $|F=1,m_{F}=0\rangle$ state and evaporatively cooled down to a
typical temperature of $\sim 1.5\mu$K in an optical
trap~\cite{Gross08}. In earlier studies of the $a$-dependent
spectrum of $K_{3}$ on this state in the vicinity of a broad
Feshbach resonance at $\sim 894$G we identified a recombination
minimum at $a^{*}_{0}= 1130(120)a_{0}$ and a maximum at
$a_{-}=-280(12)a_{0}$, where $a_{0}$ denotes the Bohr
radius\cite{Gross09,Gross11}. The ratio of these two values is close
to the universally predicted value of
$a_{-}/a^{*}_{0}=-\exp(-\pi/2s_{0})=-0.21$, where $s_{0}=1.00624$,
indicating that the three-body parameter is preserved across the
Feshbach resonance. Using universal relations we can predict the
position of the Efimov resonance at the atom-dimer threshold at
either $a_{*}\approx 1.1 a^{*}_{0}/\sqrt{22.7}=262(27)a_{0}$ or
$a_{*}\approx -1.03a_{-}=288(12)a_{0}$ using the recombination
minimum or maximum respectively~\cite{Braaten&Hammer06}.

\begin{figure}
{\centering \resizebox*{0.46\textwidth}{0.38\textheight}
{{\includegraphics{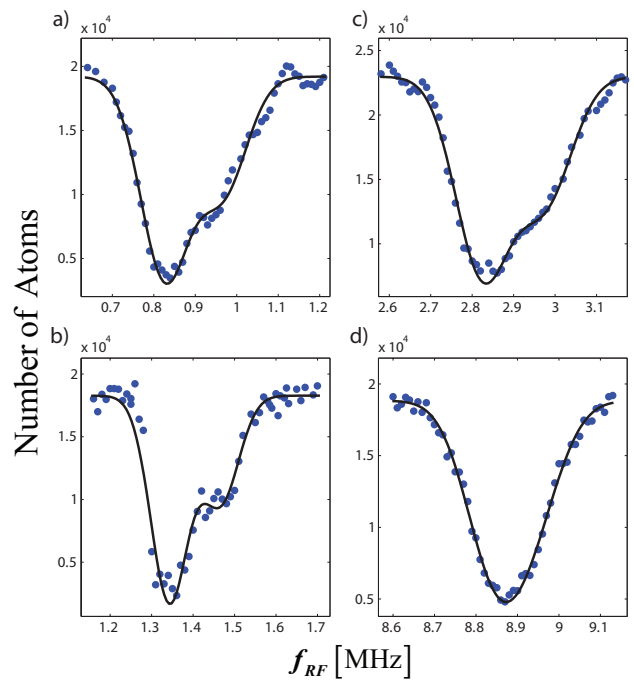}}}
\par}
\caption{\label{F2-RfScans} Rf-induced atom loss resonances are
shown for different values of the bias magnetic field corresponding
to different values of the dimer binding energy. For the weakly
bound dimer ((a)-(c)) strongly overlapped double-peak loss
structures are clearly visible. For the deeper bound dimer (d) only
a single minimum is observed, signifying that the dimer and the
trimer energy levels converged to a value below the resolution
limit. Each point is an average of 3 to 5 individual measurements.
Solid lines represent the double-peak Gaussian fits to the data.}
\end{figure}

We apply a strong rf-modulation to a given magnetic field and scan
its frequency to obtain atom loss resonances. Note that it
corresponds to the polarization of the oscillating rf-field in the
direction of the applied bias field, contrary to the orthogonal
polarization required by the spin-flip transition in
Refs.~\cite{Lompe10RF,Nakajima11}. The modulation amplitude induced
by the rf-field reaches $1.6$~G at full rf-power which allows for an
atom-dimer conversion efficiency of $\sim 40\%$ within a short time
of $1$ msec at a typical mean density of $\sim 10^{12}$~cm$^{-3}$
and $a\cong 1000a_{0}$. While the rate of molecule association is
defined by the two-body collision rate ($\sim 3$kHz for the above
parameters), the trimer association rate relies on significantly
rarer three-body collisions which require three atoms to be at the
same point in space. In the case of a dimer, quick loss is observed
when an associated dimer occasionally finds a collision partner and
decays to a deeper bound state, releasing enough kinetic energy to
cause the loss of all colliding parties. When the Efimov energy
level is populated directly, widely separated atoms that would not
necessarily collide otherwise are bound together in a weakly bound
trimer state. It is then quickly dissociated into a deeply bound
dimer and an atom, both of which leave the trap due to high kinetic
energy. Consequently, inelastic losses are resonantly enhanced which
allow us to adjust the rf-modulation time at each magnetic field
value so that it is an order of magnitude less than the
corresponding lifetime of the atoms due to the three-body
recombination. The latter is defined as $1/\tau = K_{3}\langle
n^2\rangle$, where $\langle n^2\rangle$ is the mean of the squared
atomic density~\cite{Gross09}. The rf-modulation time varies
accordingly, between $\sim 50-2000$ msec for different magnetic
fields. After switching off the rf-field we detected the remaining
number of atoms in the trap. A number of rf-scans for different
values of the magnetic field are shown in
Fig.~\ref{F2-RfScans}(a)-(d). A typical signal displays a strongly
overlapping but clearly visible double-peak loss structure in which
the lower rf-frequency and deeper loss minimum correspond to the
well known dimer association resonance, which we used earlier for
precise characterization of the Feshbach resonance~\cite{Gross11}.
However, this signal has now been power-broadened due to the high
rf-power. We associate the second loss minimum with the formation of
Efimov trimers. For the low energy dimer, the trimer signal is
clearly visible (Fig.~\ref{F2-RfScans}(a)-(c)). However, for the
deeper bound dimer it disappears (Fig.~\ref{F2-RfScans}(d)), as
expected, when the trimer and dimer energy levels' separation are
less than the typical width of the power-broadened dimer resonance
($\sim 110$kHz). Narrower dimer resonances were observed for lower
rf-powers and shorter rf-modulation times but at the expense of
losing the trimer signal. This trend can be observed in
Fig.~\ref{F2-RfScans}(b) which shows a narrower resonance structures
accompanied by a weaker trimer signal as compared to
Fig.~\ref{F2-RfScans}(a),(c). This behavior is caused by variations
in the experimental conditions of the rf-modulation pulse.

It is interesting to note that the rf-scans in Fig.~\ref{F2-RfScans}
are well fitted with two overlapping Gaussian distributions of the
same widths. Most probably this indicates that the trimer resonance
width is not defined by the finite lifetime of trimers which should have then
exceeded $10\mu$s. On the other hand, trimers do not live
longer than the inverse of the two-body collision rate which limits
their lifetime from above to several hundreds of $\mu$s for the typical
phase space densities used in the experiment. This range opens an
experimentally realistic time slot to manipulate the trimer state.
However, the direct rf-association of trimers discussed here limits
their production rate to the rate of three-body collisions which is
significantly slower than their decay rate that relies on two-body
collisions. This might prevent the observation of macroscopic
ensembles of Efimov trimers in the current approach.

\begin{figure}
{\centering \resizebox*{0.46\textwidth}{0.22\textheight}
{{\includegraphics{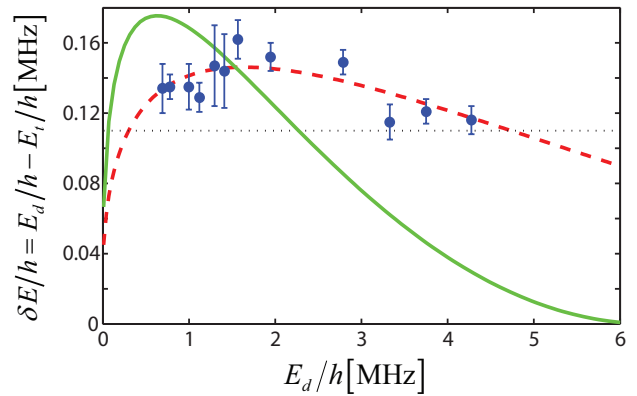}}}
\par}
\caption{\label{F3-TrimerEnergyLevel} The difference between trimer and
dimer energies as a function of the dimer energy level. The points
represent results of fits to the measurements with double-peak
Gaussian distributions. The error bars show one standard deviation
of the fit. The solid line represents the prediction made by universal theory for
$a_{*}=288a_{0}$. The dashed red line shows the fit of universal theory with
an additional fitting parameter - the multiplication factor of the
energy difference $\delta E$. The dotted horizontal line marks the
minimal $\delta E$ that can be reliably detected due to the
resolution limit.}
\end{figure}

We fit the rf-scans with a double-peak Gaussian distribution to
determine the position of the dimer resonance, its width and the
difference between the dimer and trimer energies $\delta
E=E_{d}-E_{t}$. In Fig.~\ref{F3-TrimerEnergyLevel}, $\delta E$ is
shown as a function of the dimer energy $E_{d}$. By representing
only $\delta E$ we eliminate the common shift of the energy levels
due to the finite temperature and strong magnetic field modulation. The
points in Fig.~\ref{F3-TrimerEnergyLevel} represent measurements,
while the solid green line shows the prediction made by universal theory with
$a_{*}=288a_{0}$ which is derived from the $K_3(a)$ measurements. At
first glance there appears to be a clear discrepancy between the measurements and the theory.
The universal theory predicts that the trimer and the dimer
energy levels will converge sooner than was observed in the experiment. For
example, for $E_{d}\approx h \times 2.8$MHz $\delta E$ should be
well below the resolution limit, preventing the observation of the
trimer signal, but it is clearly visible and disappears for
significantly larger $E_d$ (see Fig.~\ref{F2-RfScans}(c),(d)). Note
however, that universal theory predicts the depth of the trimer
state $\delta E$ reasonably well, leaving the main disagreement
between the theory and the experimental results in the convergence of the
two energy levels. Our measurements then suggest that the Efimov
resonance at the atom-dimer threshold is shifted to a lower
scattering length value, but they also suggest that the theory needs to go
beyond universality in order to correctly describe our data. If we were using
a lower $a_{*}$ value within the framework of universal theory then we
would obtain larger $\delta E$ in the relevant range of $E_{d}$ which
clearly contradicts our results (see
Fig.~\ref{F1-EfimovScenario}(b)). A theory of the full Efimov energy
spectrum with finite range corrections for bosonic lithium does not
yet exist. However, recently an effective field theory approach has
been used~\cite{Ji10} to calculate the shift of $a_{*}$ to a new
position $a_{*e}$ based on $R_{e}(a)$ and $K_3(a)$ which were
reported in our previous work~\cite{Gross09}. This ``beyond
universality'' theory predicts that $a_{*e}=210(44)a_{0}$ in agreement
with the general trend of the experimental results. In the absence
of a full theory and in order to estimate a value of $a_{*e}$ with which the
measurements represented in Fig.~\ref{F3-TrimerEnergyLevel} will
agree, we modify the universal theory by introducing an additional
parameter, a multiplication factor of the energy difference $\delta
E$. We fit our data with this model and the result is shown in
Fig.~\ref{F3-TrimerEnergyLevel} as a dashed red line which describes
our measurements reasonably well and roughly estimates
$a_{*e}\approx 180a_{0}$. Let us stress, however, that it is necessary to fit
our data with a complete theory in order
to correctly predict the value of $a_{*e}$.

\begin{figure}
{\centering \resizebox*{0.46\textwidth}{0.22\textheight}
{{\includegraphics{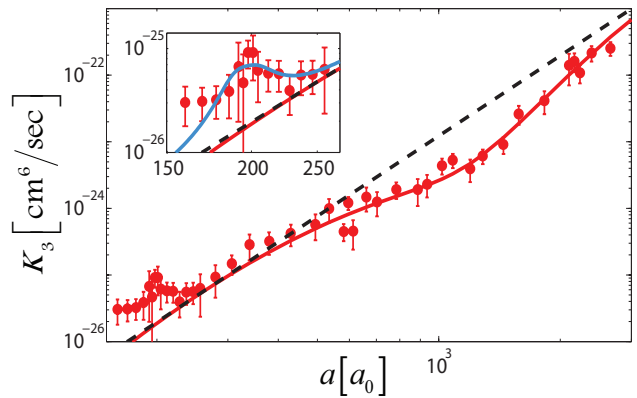}}}
\par}
\caption{\label{F4-K3} The three-body recombination spectrum. Only
the positive range of $a$ is shown. In addition to the earlier published
minimum in $K_{3}(a)$\cite{Gross10} there is a local maximum
(enlarged in the inset) that we attribute to a secondary collision
resonance which indicates the position of the Efimov resonance at
the atom-dimer threshold $a^{*}_{e}\approx 200a_{0}$. The solid line
in the inset is a fit to the model of secondary
collisions~\cite{Machtey11a}.}
\end{figure}

Independently, $a_{*e}$ can be measured via the resonance from secondary collisions
which is expected to appear in the $K_3(a)$ spectrum~\cite{Zaccanti09}.
This resonance occurs when a dimer with relatively
high kinetic energy is formed in a three-body recombination process
and collides with other atoms on its way out of the trap, resulting
in the loss of more than three atoms in a single three-body
recombination event. Such resonance is likely to occur in the
vicinity of the Efimov resonance at the atom-dimer threshold where
atom-dimer elastic and inelastic collisional cross-sections are
enhanced by orders of magnitudes~\cite{Braaten&Hammer06}. Here we
reveal the $K_3(a)$ measurement that we performed earlier on the
absolute ground state ($|F=1,m_F=1\rangle$)~\cite{Gross10,Gross11}.
Fig.~\ref{F4-K3} shows $K_3(a)$ for positive scattering lengths in
the vicinity of the Feshbach resonance at $\sim 738$G. In addition
to the already indicated three-body recombination minimum at
$a^{*}_{0}=1260(76)a_{0}$, a narrow peak appears at $\sim 200a_{0}$.
We can now attribute this peak to a secondary collision resonance
that reveals the position of $a_{*e}$. To make a quantitative
argument in favor of this attribution we developed a model which
counts the elastic and inelastic collisions of a dimer with trapped
atoms based on the available analytical expressions for the
cross-sections of these events~\cite{Machtey11a}. Thus the model
predicts enhancement of atom loss per three-body recombination
event. We fit our data with this model using three fitting
parameters: the height of the peak (which has no meaning as we do not
model the $K_{3}$ coefficient directly), $a_{*e}$, and $\eta$ which
describes the lifetime of a trimer. The fit, shown as a solid line
in the inset of Fig.~\ref{F4-K3}, shows good agreement with
the measurement. This supports our treatment of the feature as an
avalanche resonance. We note that the fitting parameter
$\eta=0.072(1)$ is smaller by a factor of $\sim 2.5$ compared to the
experimentally observed values for the minimum and maximum in the
$K_{3}(a)$ spectrum\cite{Gross10,Gross11}. This disagreement might
be attributed to yet another manifestation of the ``beyond
universality'' corrections required to explain our experimental
results.

The fitted position of $a_{*e}=196(4)a_{0}$ does not agree with the
universal theory's prediction of either $a_{*}=292(18)a_{0}$ or
$a_{*}=282(12)a_{0}$ using $a^{*}_{0}$ or $a_{-}$, respectively.
However, it does agree with the general trend of the rf-association
results performed on a different state, which predict the shift of
$a_{*}$ to lower values as well. Remarkably, the measured position
of $a_{*e}$ in the absolute ground state corresponds well to the
value predicted by the ``beyond universality'' theory for the second
to lowest Zeeman sublevel~\cite{Ji10}. This similarity may not be
surprising since the other Efimov features were discovered to be
identical for both levels within the experimental
errors~\cite{Gross10,Gross11}. Moreover, it was shown that the
effective range $R_{e}$ and the parameters of the Feshbach
resonances on both levels differ
insignificantly~\cite{Gross09,Ji10,Gross11}. However, as we noted
earlier, the fitting of our rf-association measurements to a
complete theory will allow a quantitative comparison between the
values of $a_{*e}$ and, thus, a detailed analysis of the influence
of the finite range corrections on both levels. We note, finally,
that our original $K_3(a)$ measurements on the $|F=1,m_{F}=0\rangle$
level~\cite{Gross09} were not dense enough in the relevant region of
$a$ to allow the identification of a similar avalanche resonance and
they have to be revisited to reveal this feature.

It is interesting to compare our results with those obtained for
$^6$Li, especially in light of the experimentally observed
universality in the position of the Efimov resonance at the free
atom threshold ($a_{-}$) in the region of $a<0$. In all but one
atomic species $a_{-}/r_{\textrm{vdW}}\sim -9.5$ within $\pm
15\%$~\cite{Berninger11,Zaccanti09,Pollack09,Gross09,Gross10,Ottenstein08,Lompe10,Huckans09,Williams09,Wild11}.
This was recently explained in a landmark theoretical
paper~\cite{Wang11,footnote02}. Despite this universality in
$a_{-}$, $a_{*e}$ varies significantly between different
species~\cite{Lompe10,Nakajima10,Knoop09,Zaccanti09,Machtey11a}
which is a clear consequence of corrections caused by the finite
range of interatomic potentials. However, though $r_{\textrm{vdW}}$ for
both isotopes of Li is very similar~\cite{Chin10}, $a_{*e}$
still differs significantly. In our case $a_{*}/a_{*e}\sim 1.5$, but for
$^6$Li corrections go in the opposite direction, i.e.
$a_{*}/a_{*e}<1$~\cite{Lompe10,Nakajima10}. This emphasizes the role
played by the $s_{res}$ parameter in the breakdown of three-body
universality in $^7$Li.

In conclusion, we developed a general experimental technique which paves the
way for direct population, manipulation and interrogation of Efimov
quantum states. During the short but experimentally reasonable lifetime
of Efimov trimers, they can be moved into a deeply bound state
and their macroscopic ensemble can then be realized. We also note
that a recent theory has found reasonable rf-association
rates of Efimov trimers in $^7$Li for similar phase space
densities~\cite{Tscherbul11}. However, it uses a different
starting configuration (one boson is distinguishable) and thus
cannot be directly applied to our system.

LK acknowledge stimulating discussions with L.~Platter,
E.A.~Cornell, E.~Braaten and S.~Jochim. We thank
S.J.J.M.F.~Kokkelmans for many inspiring discussions and critical
reading of the manuscript. This work was supported, in part, by the
Israel Science Foundation.


\begin{thebibliography}{36}
\bibitem{footnote01}
\bibinfo{note}{Earlier theoretical treatments, especially within the framework of nuclear physics, are reviewed in Ref.~\cite{Braaten&Hammer06}}.
\end{thebibliography}

\end{document}